\documentclass{optica-article}
\articletype{Research Article}
\usepackage{amsmath}
\usepackage{lineno}

\makeatletter
\providecommand{\@journalname}{Optics Express}
\providecommand{\@articletype}{Research Article}
\makeatother
\begin{document}

\title{Efficient imaging of quantum emitters using compressive sensing}

\author{Sonali Gupta,\authormark{1} Kiran Bajar,
\authormark{1} Alexander McFarland,\authormark{2} Amit Kumar,\authormark{1} Subhas Manna,\authormark{1} and Sushil Mujumdar,\authormark{1,*}}

\address{\authormark{1}Tata Institute of Fundamental Research, Dr. Homi Bhabha Road, Colaba, Mumbai, 400005, Maharashtra, India}

\address{\authormark{2}Department of Electrical Engineering and Computer Science, University of Michigan, Ann Arbor, MI, 48109, USA}

\email{\authormark{*}mujumdar@tifr.res.in}

\begin{abstract}

Optical imaging of quantum emitters is essential for a wide range of quantum applications. Conventional confocal imaging relies on point-by-point raster scanning, which is inherently time-consuming and photon-inefficient, particularly for sparse emitter distributions and photon-limited samples. Here, we demonstrate a compressive sensing-based imaging approach, where spatially structured wide-field excitation replaces raster scanning, enabling reconstruction of sparse emitter. In our implementation, random binary patterns are used to acquire compressive measurements, from which the spatial fluorescence distribution is reconstructed using a GPSR-BB algorithm. We experimentally demonstrate this approach using nitrogen-vacancy (NV) centers in diamond as a representative platform, with high-fidelity image reconstruction achieved using only approximately $20\%$ of the measurements required for conventional raster scanning. In addition to intensity reconstruction, we extend this framework to reconstruct spatial maps of the second-order correlation function $g^{(2)}(0)$ from compressive measurements. This enables identification of single-photon emitters through antibunching signatures using significantly reduced data.
Accurate reconstruction is achieved despite strong undersampling, demonstrating reduced measurement overhead while preserving spatial information and enabling correlation-based quantum imaging.
\end{abstract}

\section{Introduction}

Optical imaging of quantum emitters is of fundamental importance across a wide range of scientific and technological applications, including quantum photonics, nanoscale sensing, super-resolution microscopy, and quantum information processing \cite{Aharonovich2016, Rondin2014, Hell2007}. Spatially resolved measurements of individual emitters enable precise characterization of their optical properties, spatial distributions, and interactions with the local environment, which are essential for the development of scalable quantum technologies \cite{Awschalom2018,Sutula2023}. Recent advances in quantum technologies increasingly demand imaging techniques that can operate over large fields of view, under photon-limited conditions, and with minimal acquisition time, motivating the exploration of alternative imaging strategies beyond conventional scanning-based approaches \cite{Casola2018, Genovese2016}. In this context, adaptive optics is emerging as an important tool in quantum imaging, particularly under low-photon conditions \cite{Bajar2026,Bajar2025}.

Optical imaging of quantum emitters is conventionally performed using confocal microscopy, where a tightly focused excitation spot is raster scanned point by point across the sample using piezoelectric stages to construct the full image \cite{Wilson1994,Gruber1997}. Although this approach provides high spatial resolution and efficient background rejection, it is inherently time-consuming, as the sample must be sequentially excited and measured at every spatial pixel. This limitation becomes particularly restrictive when imaging large fields of view, weak emitters, or samples with limited photon budgets. Moreover, raster scanning treats all spatial locations uniformly, leading to significant measurement overhead in regions that do not contain emitters \cite{Studer2012,Wolf2015}. This inefficiency is especially pronounced for sparse quantum systems, where only a small fraction of the field of view contributes meaningful signal.

To overcome these constraints, we introduce an imaging approach based on compressive sensing, in which the sample is illuminated using wide-field, spatially structured excitation patterns rather than point-by-point scanning \cite{Duarte2008, Wang:23}. Unlike raster-scanned imaging, compressive sensing relies on multiplexed measurements that simultaneously probe multiple spatial locations, thereby increasing the information content of each measurement \cite{Candes2008,Edgar2018}. The image is subsequently computationally reconstructed from a significantly reduced number of measurements, enabling image acquisition in a fraction of the time required by conventional raster scanning. This technique exploits the fact that many quantum emitter distributions are intrinsically sparse in real space, allowing accurate image recovery from undersampled measurements. As a result, compressive sensing provides a photon-efficient and time-efficient alternative for imaging sparse quantum systems. The second-order correlation function $g^{(2)}(0)$ provides information about the emitter number and is widely used to identify single-photon sources via antibunching \cite{Brouri:00,Lounis2000, Kurtsiefer2000}. Extending compressive sensing to reconstruct such correlation-based quantities provides richer information while retaining reduced acquisition time and photon efficiency.

In this work, we use NV centers in diamond as a representative solid-state quantum emitter platform \cite{Liu2018}; however, the imaging approach presented here is general and applicable to a broad class of quantum emitters \cite{Michler2000, Tran2015, Flatae2024}, whose utility requires sparse spatial distributions. NV centers provide an ideal test system due to their localized emission, sparse spatial distribution, and well-characterized optical properties, enabling direct comparison between compressive-sensing-based reconstructions and conventional raster-scanned images \cite{Wrachtrup2006}. We employ a digital micromirror device (DMD) to project a sequence of random binary masks onto the sample, enabling spatially structured wide-field excitation \cite{Duarte2008}. For each applied mask, the total fluorescence emitted by the sample is collected using a single-photon avalanche diode (SPAD), resulting in a set of compressive measurements. The acquired data are then used to reconstruct the spatial fluorescence distribution using a gradient projection for sparse reconstruction algorithm with Barzilai--Borwein step size (GPSR-BB) \cite{Figueiredo2007,BARZILAI1988,10.5555/2526243}. Using this approach, we demonstrate accurate image reconstruction using only $20\%$ of the full measurement data. This enables identification of potential single-photon emitters within the field of view. We numerically extend this framework to reconstruct spatial maps of the second-order correlation function $g^{(2)}(0)$ using compressive measurements and converge onto actual single-photon emitter locations. 
This highlights the potential of compressive sensing for correlation-based quantum imaging.

Finally, the quality of the reconstructed images is quantitatively evaluated by comparison with corresponding raster-scanned reference images. We employ the peak signal-to-noise ratio (PSNR) and the normalized correlation coefficient as quantitative metrics to assess reconstruction fidelity \cite{Luo2018,OliaeiMoghadam2025}. These measures provide complementary information on the similarity between the reconstructed and reference images in terms of both intensity accuracy and spatial agreement. The obtained PSNR and correlation values indicate good agreement between the compressive sensing based reconstructions and raster-scanned images, consistent with values reported in the literature \cite{Nan2024,Yu2019,Koutsourakis2019}. For the correlation-based reconstructions, we evaluate identification of single-photon emitters using the condition $g^{(2)}(0) < 0.5$, demonstrating reliable recovery under compressive sampling.

\section{Compressive Sensing Reconstruction of Sparse Matrices}
\label{alg}

Let $\mathbf{X} \in \mathbb{R}^{N \times N}$ represents the two-dimensional image to be reconstructed. We assume that $\mathbf{X}$ admits a sparse representation in a known sparsifying transform $\boldsymbol{\Psi}$, i.e.,
\begin{equation}
\mathbf{X} = \boldsymbol{\Psi}\mathbf{S},
\end{equation}
where $\mathbf{S}\in\mathbb{R}^{N\times N}$ is a coefficient matrix with only a few non-zero entries. The sparsity level is
\begin{equation}
K = \|\mathbf{S}\|_0, \qquad K \ll N^2,
\end{equation}
where $\|\cdot\|_0$ denotes the number of non-zero elements.

For computational convenience, we use vectorized forms of the image and coefficient matrices:
\begin{equation}
\mathbf{x} = \operatorname{vec}(\mathbf{X}) \in \mathbb{R}^{N^2}, 
\qquad
\mathbf{s} = \operatorname{vec}(\mathbf{S}) \in \mathbb{R}^{N^2}.
\end{equation}
With the corresponding matrix representation of the sparsifying transform, the model becomes
\begin{equation}
\mathbf{x} = \mathbf{\Psi}\mathbf{s},
\end{equation}
where $\mathbf{\Psi}\in\mathbb{R}^{N^2\times N^2}$ is the vectorized form of the sparsifying operator (e.g., $\mathbf{\Psi}=\mathbf{I}$ for sparsity in the canonical pixel basis).

Instead of measuring each pixel of $\mathbf{X}$ individually, we acquire $M$ linear measurements:
\begin{equation}
\mathbf{y} = \boldsymbol{\Phi}\mathbf{x} + \mathbf{n},
\end{equation}
where $\mathbf{y}\in\mathbb{R}^{M}$ is the measurement vector, $\boldsymbol{\Phi}\in\mathbb{R}^{M\times N^2}$ is the sensing matrix, and $\mathbf{n}$ accounts for measurement noise (e.g., photon shot noise and detector noise). Each row of $\boldsymbol{\Phi}$ represents a DMD illumination pattern, while the corresponding element of $\mathbf{y}$ denotes the total fluorescence collected under that pattern. We operate in the compressive regime, $M \ll N^2$. In general, stable recovery of a $K$-sparse signal is possible when \cite{10.5555/1873601.1873696}
\begin{equation}
M \ge \,K \log\!\left(\frac{N^2}{K}\right),
\end{equation}

Combining the sparse model with the measurement model yields
\begin{equation}
\label{cs}
\mathbf{y} = \mathbf{A}\mathbf{s} + \mathbf{n}, 
\qquad \text{with } \mathbf{A}=\boldsymbol{\Phi}\mathbf{\Psi}\in\mathbb{R}^{M\times N^2}.
\end{equation}
Reconstruction is performed by estimating $\mathbf{s}$ from $\mathbf{y}$ using an $\ell_1$-regularized least-squares formulation \cite{Tibshirani1996}:
\begin{equation}
\hat{\mathbf{s}} = \arg\min_{\mathbf{s}} \; \frac{1}{2}\|\mathbf{y}-\mathbf{A}\mathbf{s}\|_2^2 + \tau \|\mathbf{s}\|_1,
\end{equation}
where $\tau>0$ controls the trade-off between data fidelity and sparsity. Following standard practice, we set
\begin{equation}
\label{tau}
\tau = \alpha \, \|\mathbf{A}^T \mathbf{y}\|_{\infty},
\end{equation}
with $\alpha$ typically in the range $0.1\%$--$1\%$.

Finally, the reconstructed image is obtained by reshaping the recovered vector $\hat{\mathbf{x}}=\mathbf{\Psi}\hat{\mathbf{s}}$ back into matrix form:
\begin{equation}
\hat{\mathbf{x}} = \mathbf{\Psi}\hat{\mathbf{s}}, 
\qquad
\hat{\mathbf{X}} = \operatorname{unvec}(\hat{\mathbf{x}})\in\mathbb{R}^{N\times N}.
\end{equation}

The above optimization problem is solved using the GPSR-BB, which is well suited for large-scale sparse recovery. 

In GPSR-BB, the solution vector is decomposed into positive and negative components, allowing the $\ell_1$-norm to be handled efficiently through non-negativity constraints. Starting from an initial estimate (typically a zero vector or the matched filter $\mathbf{A}^T\mathbf{y}$), the algorithm iteratively updates the solution by taking gradient descent steps on the quadratic data-fidelity term, followed by projection onto the feasible set.

To improve convergence and numerical stability, a continuation strategy is employed in which the regularization parameter $\tau$ is gradually reduced. An optional debiasing step can be applied at the final stage to mitigate the bias introduced by the $\ell_1$ penalty and refine the reconstruction. Convergence is monitored through relative changes in the objective function and the solution vector.

In the present case, sparsity is assumed in the canonical pixel basis ($\mathbf{\Psi} = \mathbf{I}$), and therefore the effective sensing matrix reduces to $\mathbf{A} = \boldsymbol{\Phi}$, yielding the measurement model $\mathbf{y} = \mathbf{A}\mathbf{x}$. Experimentally, the NV-center sample is illuminated using a sequence of random binary patterns generated by the DMD, and for each pattern the total fluorescence signal is collected using a single-pixel detector. These measurements directly implement the linear sensing model introduced above, where each DMD pattern forms one row of the sensing matrix $\mathbf{A}$, and the corresponding measured fluorescence intensity forms one element of the measurement vector $\mathbf{y}$.


\section{Numerical simulations}

To validate the proposed compressive sensing framework, we perform numerical simulations of sparse NV center distributions. A two-dimensional matrix representing the sample plane is generated, where emitter locations are randomly distributed across the field of view. Each occupied pixel contains between 1 and 4 NV centers, resulting in a spatially sparse intensity distribution.

To simulate fluorescence emission from each NV center, we model the emitter as a three-level system consisting of ground (G), excited (E), and metastable shelving (S) states \cite{GUPTA2026101004}. Optical excitation drives transitions $G \rightarrow E$ at rate $k_{ge}$. The excited state decays radiatively ($E \rightarrow G$) at rate $\gamma$, producing detectable photons, or non-radiatively via intersystem crossing ($E \rightarrow S$) at rate $k_{es}$. Population in the shelving state returns to the ground state at rate $k_{sg}$. Photon emission events correspond to radiative $E \rightarrow G$ transitions. Simulating these transitions in time gives the photon arrival timestamps for each emitter.

\begin{figure}[ht]
\centering
\includegraphics[scale=0.45]{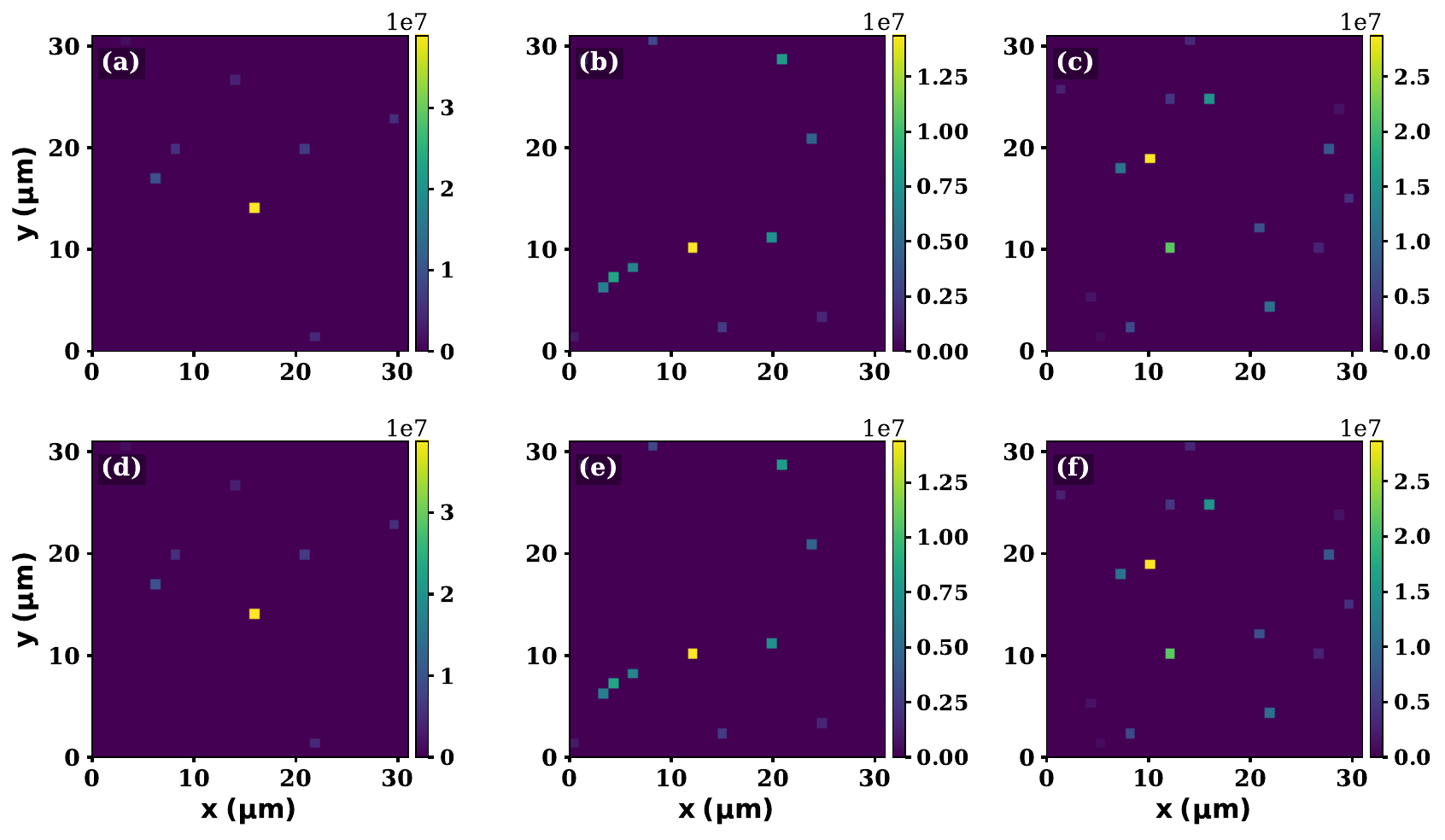}
\caption{Simulation result. Comparison of ground-truth and reconstructed images. Panels (a)-(c) show the ground-truth images, while panels (d)-(f) show the corresponding images reconstructed using the compressive sensing algorithm with only \(20\%\) of the measurement data. From left to right, the sparsity decreases, corresponding to scenes with 8, 12, and 16 occupied pixels, respectively.}
\label{simulations}
\end{figure}

To account for realistic excitation, the illumination is modeled with a Gaussian spatial profile. Each pixel $j$ is weighted as
\begin{equation}
w_j = \exp\!\left(-\frac{d_j^2}{2\sigma^2}\right),
\end{equation}
where $d_j$ is the distance from the beam center and $\sigma$ denotes the beam width. Photon timestamps from pixel $j$ are retained with probability $p_j = w_j$. Subsequently, realistic detection effects are applied independently to each channel, including a detector dead time of $55\,\mathrm{ns}$, background fluorescence contributing approximately $10\%$ of the detected counts per channel, detector timing jitter of $400\,\mathrm{ps}$, and a finite detection efficiency of $80\%$. The resulting weighted and noise-affected photon counts define the pixel intensities, yielding the sparse spatial distribution shown in Fig~\ref{simulations}(a-c).

Here, $\mathbf{A}$ is formed by the random DMD patterns, while $\mathbf{y}$ contains the corresponding total fluorescence measurements. The pixel intensity distribution $\mathbf{x}$ is then reconstructed using the compressive sensing approach described in Section \ref{alg}.

Fig~\ref{simulations}(d-f) shows the reconstructed image obtained using only $20\%$ of the total measurement data. Despite strong undersampling, the reconstructed image accurately reproduces both the spatial locations and the relative intensities of the NV centers. To quantify reconstruction fidelity, we compute the normalized correlation coefficient between the ground-truth image and the reconstructed image, which is found to be $100\%$ under the present simulation conditions. This demonstrates that, for sufficiently sparse emitter distributions and realistic excitation modeling, the compressive sensing approach enables accurate recovery of the NV center intensity map even when a substantial fraction of the measurements are discarded.

\section{Experiment}
\begin{figure}[ht]
\centering
\includegraphics[scale=0.75]{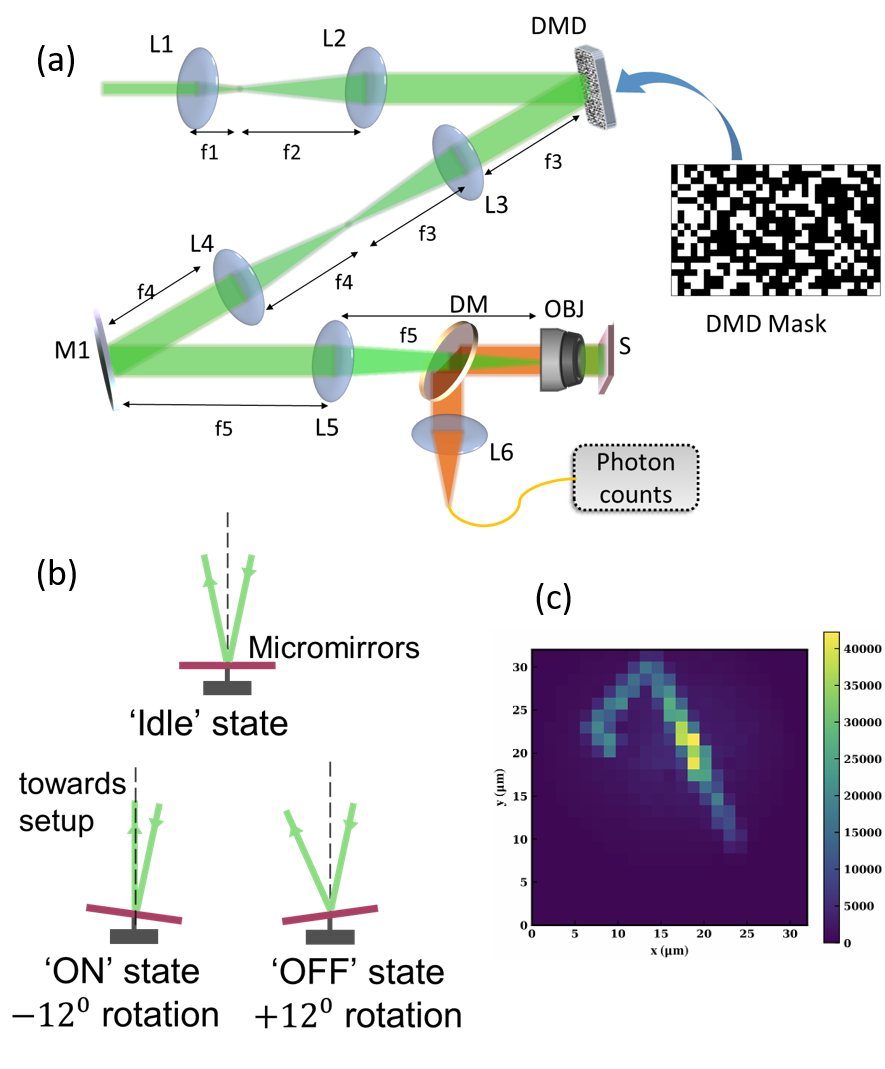}
\caption{(a) Schematic of the experimental setup for compressive sensing imaging using a digital micromirror device (DMD). \(L_i\) denotes lenses with focal lengths \(f_i\). The excitation laser is expanded using \(L_1\) and \(L_2\) and directed onto the DMD for spatial modulation. For pattern generation, \(10 \times 10\) DMD micromirrors are grouped to form one superpixel. The inset shows a representative binary DMD mask, where white and black correspond to ON and OFF superpixels, respectively and \(40\%\) of the superpixels are set to the ON state. Using lenses \(L_3\) and \(L_4\), the DMD plane is imaged onto mirror \(M_1\). The patterned light reflected from \(M_1\) is then relayed by lens \(L_5\) and a \(100\times\) microscope objective (OBJ) onto the sample (S). A dichroic mirror (DM) with a cutoff wavelength of \(560\,\mathrm{nm}\) transmits the excitation light and reflects the longer-wavelength fluorescence toward the detection arm. The emitted photons are collected and recorded as photon counts for image reconstruction. (b) Schematic illustration of the two switching states of a digital micromirror device (DMD). Each micromirror can tilt by \(\pm 12^\circ\) with respect to its idle position. In the ON state \((-12^\circ)\), the reflected light is directed toward the optical setup, whereas in the OFF state \((+12^\circ)\), the reflected light is directed away from the setup. (c) Raster scanned image of the USAF resolution target, to verify that the collection fiber collects signal from all relevant spatial modes in the system. The clear recovery of the target features indicates that the detection path provides sufficient spatial-mode collection.}
\label{Expsetup}
\end{figure}

The experimental setup is illustrated schematically in Fig~\ref{Expsetup}(a). A continuous-wave, fiber-coupled laser operating at a wavelength of $532\,\mathrm{nm}$ is used as the excitation source. The output beam from the fiber has an initial diameter of approximately $2\,\mathrm{mm}$, which is expanded to $6\,\mathrm{mm}$ using a $3\times$ beam magnification telescope. The expanded beam is then incident on a DMD, which consists of an array of $1280 \times 800$ micro-mirrors, each with a physical size of $10.8 \times 10.8\,\mu\mathrm{m}^2$. Each micro-mirror can be individually tilted to either $+12^{\circ}$ or $-12^{\circ}$ with respect to the DMD surface as shown in Fig~\ref{Expsetup}(b) . Each micromirror deflects the reflected beam by twice its tilt angle with respect to the incident beam, giving a total angular separation of \(24^\circ\) between the two reflection states. Mirrors oriented at $-12^{\circ}$ direct the incident light along the optical path toward the sample and therefore labeled as bright pixels, while mirrors oriented at $+12^{\circ}$ deflect the light away from the collection path and labeled as dark pixels. By applying spatially varying binary patterns consisting of bright and dark pixels, the DMD modulates the incident wavefront.

The modulated beam is relayed to the sample using a cascaded $4f$ imaging system (\(L_3\), \(L_4\), \(f_3\) = \(f_4\)). The $4f$ configuration images the DMD plane onto the mirror M1, which is placed at the back focal plane of a lens \(L_5\) with focal length $f_5 = 20\,\mathrm{cm}$. The lens \(L_5\) along with the microscope objective ($100\times$) forms a second $4f$ system, resulting in projecting the image onto the sample with an overall demagnification factor of approximately $100\times$. Consequently, the $6\,\mathrm{mm}$ beam incident on the DMD is reduced to an illumination area of approximately $60\,\mu\mathrm{m}$ at the sample plane.The sample consisted of nanodiamonds containing NV centers drop-cast onto a silicon wafer substrate, resulting in a spatially sparse fluorescence distribution well suited for compressive sensing recovery.

The same microscope objective is used to collect the fluorescence emitted by the NV centers. The collected fluorescence is coupled into a multimode optical fiber, enabling efficient collection of multiple spatial modes, and is subsequently detected using a single-photon avalanche diode (SPAD). During the measurement, a sequence of random binary masks is displayed on the DMD as shown in Fig~\ref{Expsetup}(a). The native DMD pixels, each of size are grouped into $10\times10$ super-pixels. Upon projection onto the sample through a $100\times$ microscope objective, each super-pixel corresponds to an effective illumination area of approximately $1\,\mu\mathrm{m}$ on the sample plane. Each mask is activated for an integration time of $T = 1\,\mathrm{s}$. Each activated mask produces a distinct spatial excitation pattern and corresponds to one row of the sensing matrix $\mathbf{A}$ defined in Eq.~\ref{cs}. For each applied mask, the corresponding total fluorescence signal emitted by the NV centers is recorded by the SPAD, forming an element of the measurement vector $\mathbf{y}$ in Eq.~\ref{cs}. Using the sensing matrix $\mathbf{A}$ and the measurement vector $\mathbf{y}$, the regularization parameter $\tau$ is chosen according to Eq.~\ref{tau}, and the image is reconstructed using the sparse recovery algorithm described in Section~\ref{alg}. Using this approach, accurate image reconstruction is achieved using only approximately $20\%$ of the measurements required for conventional raster scanning.


To validate the performance of the imaging platform, a high resolution USAF resolution target was used as a calibration sample. The target was imaged using DMD-based raster scanning while reflected was collected through the multimode fiber detection path. The accurate reconstruction of the known spatial features of the resolution chart confirms that the detection system efficiently collects the emitted fluorescence without spatial mode selectivity and faithfully reproduces the ground-truth structure Fig~\ref{Expsetup}(c). This validation establishes the reliability of the setup for both raster-scanned and compressive sensing measurements.

\begin{figure}[ht]
\centering
\includegraphics[scale=0.5]{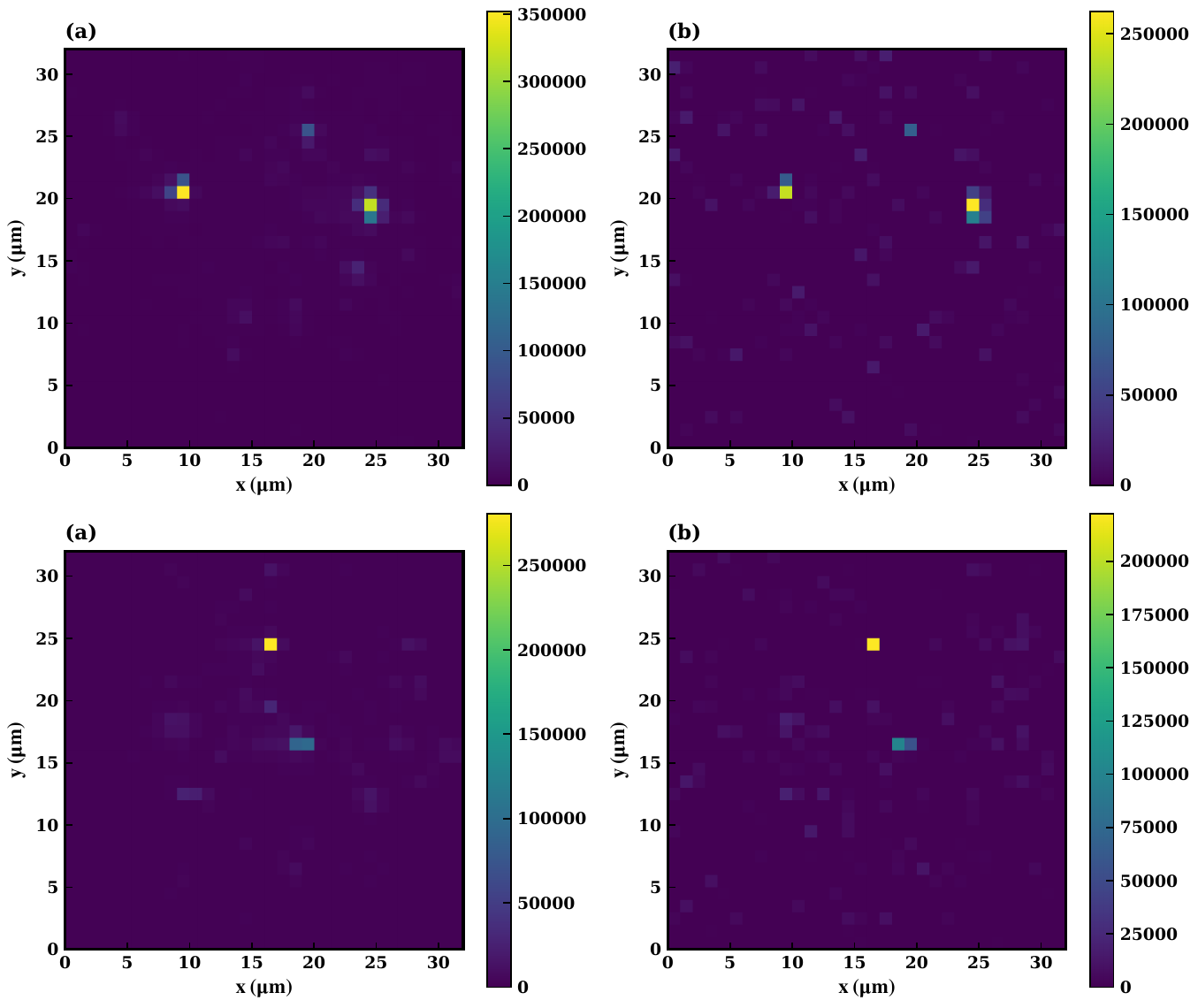}
\caption{Experimental comparison of raster scanning and compressive sensing reconstruction. 
(a) Reference image obtained by DMD-based raster scanning. 
(b) Image reconstructed using the proposed compressive sensing algorithm from $20\%$ of the measurements. 
The reconstructed image shows strong agreement with the reference, with a correlation coefficient of $0.927$.}
\label{Recimg}
\end{figure}

For the reference image shown in Fig~\ref{Recimg}(a),(c), a conventional raster-scanning procedure was implemented by sequentially activating a single DMD super-pixel while keeping all others in the off state. For an image comprising $n \times n$ spatial elements, this approach requires $n^2$ independent measurements, resulting in a correspondingly long acquisition time. In contrast, the compressive sensing reconstruction presented in Fig~\ref{Recimg}(b),(d) was obtained using a sequence of random binary masks. In each mask, approximately $40\%$ of the super-pixels were simultaneously activated, leading to spatially multiplexed excitation. Each measurement therefore encodes information from multiple spatial locations, increasing the information content per acquisition. Leveraging the sparsity of the NV fluorescence distribution, the full $n \times n$ image was reconstructed using only $0.2\,n^2$ measurements, corresponding to $20\%$ of the raster-scan dataset. This results in an effective fivefold reduction in acquisition time while maintaining high reconstruction accuracy.

\begin{figure}[ht]
\centering
\includegraphics[scale=0.7]{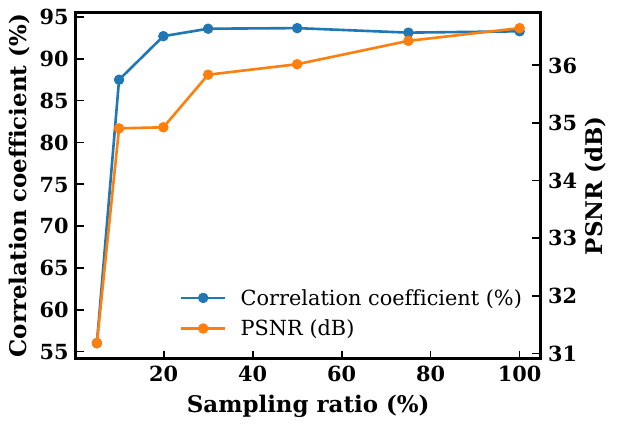}
\caption{
Correlation coefficient and PSNR versus sampling ratio, demonstrating improved reconstruction fidelity with increasing measurement fraction and saturation at higher sampling levels.
}
\label{psnr_corr}
\end{figure}

Reconstruction fidelity was quantified using the Pearson correlation coefficient and the peak signal-to-noise ratio (PSNR). For a sampling ratio of $20\%$, the reconstructed image exhibits a correlation of $0.927$ with respect to the raster-scanned reference and a PSNR of $35\,\mathrm{dB}$. The dependence of these metrics on sampling ratio is shown in Fig~\ref{psnr_corr}, where both quantities increase rapidly with measurement fraction and approach saturation beyond approximately $30\%$. These results demonstrate reliable image recovery under substantial undersampling.

\section{Compressive sensing scheme for spatial \(g^{(2)}(0)\) reconstruction}

In the previous section, we experimentally demonstrated compressive sensing reconstruction of the fluorescence intensity image using only \(20\%\) of the measurements. We now numerically demonstrate an extension of this framework to reconstruct the spatial map of the second-order correlation function \(g^{(2)}(0)\). Unlike intensity imaging, however, the measured correlation corresponding to a compressive measurement is not given by a simple linear sum of the pixelwise correlations. A modified reconstruction formalism is therefore required.

Consider a single compressive measurement in which fluorescence from multiple pixels is simultaneously collected. Let \(I_i\) denote the intensity contribution from the \(i\)-th pixel and \(g_i^{(2)}(0)\) its corresponding second-order correlation value. The measured correlation of the aggregated signal is then given by
\begin{equation}
g^{(2)}_{\mathrm{tot}}(0)=
\frac{\sum_i g_i^{(2)}(0) I_i^2 + \sum_{i\neq j} I_i I_j}
{\left(\sum_i I_i\right)^2}.
\end{equation}
Using \(\left(\sum_i I_i\right)^2 = \sum_i I_i^2 + \sum_{i\neq j} I_i I_j\), the above expression can be rearranged as
\begin{equation}
\left[1-g^{(2)}_{\mathrm{tot}}(0)\right]
\left(\sum_i I_i\right)^2
=
\sum_i
\left[1-g_i^{(2)}(0)\right] I_i^2 .
\end{equation}

For each DMD illumination pattern, fluorescence photon timestamps are recorded and used to calculate both the total intensity \(I_{\mathrm{tot},k}\) and the corresponding second-order correlation value \(g^{(2)}_{\mathrm{tot},k}(0)\) for the \(k\)-th compressive measurement. We therefore define the measurement vector as
\begin{equation}
y_k =
\left[1-g^{(2)}_{\mathrm{tot},k}(0)\right]
\left(I_{\mathrm{tot},k}\right)^2 .
\end{equation}

The sensing matrix is given by \(A_{k,i}\) where \(A_{k,i}\) denote the binary value of the DMD pattern at pixel $i$ for the \(k\)-th measurement. We define the unknown quantity
\begin{equation}
x_i = \left[1-g_i^{(2)}(0)\right] I_i^2 .
\end{equation}
The forward model for the compressive measurements can then be written as
\begin{equation}
y_k = \sum_i A_{k,i} x_i,
\end{equation}
or, equivalently, in matrix form,
\begin{equation}
\mathbf{y} = A \mathbf{x}.
\end{equation}

After solving this inverse problem using the same reconstruction algorithm employed for the intensity image, the spatial values of \(g_i^{(2)}(0)\) are recovered using the intensity map obtained in the first stage:
\begin{equation}
g_i^{(2)}(0) = 1 - \frac{x_i}{I_i^2}.
\end{equation}
Here, \(I_i\) denotes the intensity of the \(i\)-th pixel reconstructed from the compressive intensity measurements. This leads to a two-step reconstruction procedure: first, the intensity map is recovered from the compressive fluorescence measurements; second, this intensity information is used together with the compressively measured correlation data to reconstruct the spatial \(g^{(2)}(0)\) map. In this way, the compressive sensing framework is extended from intensity imaging to correlation-based imaging.

\begin{figure}[ht]
\centering
\includegraphics[scale=0.5]{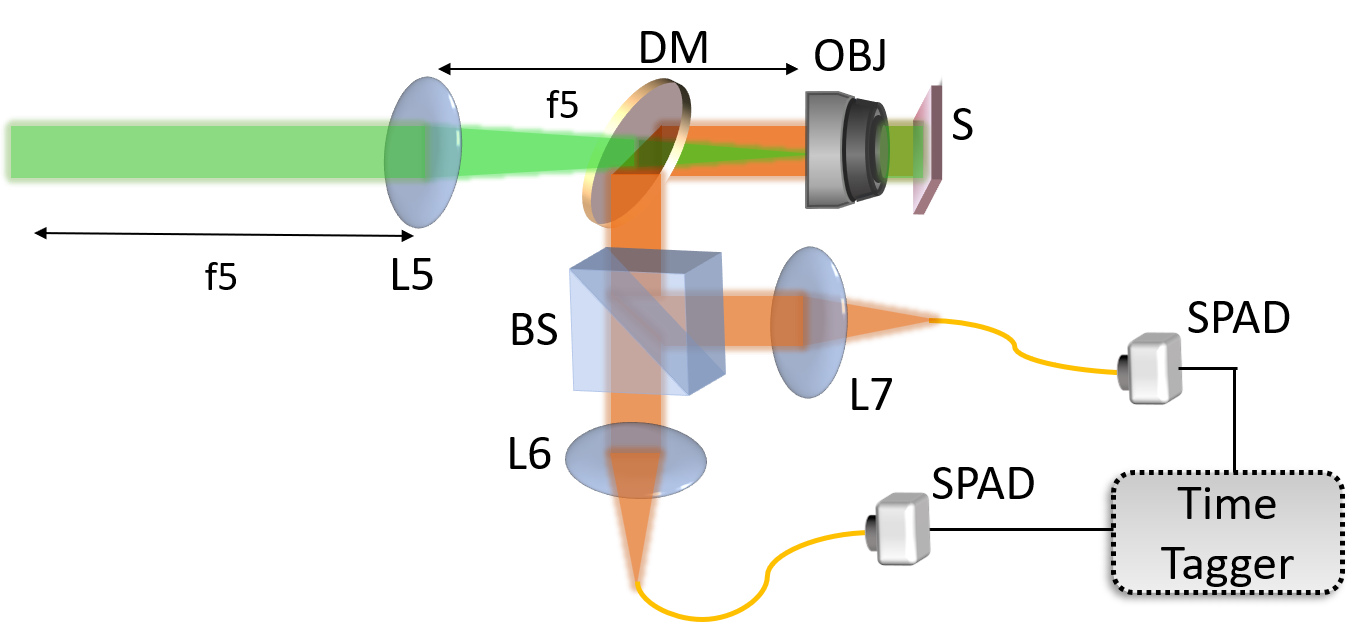}
\caption{Modified version of the setup shown in Fig.~\ref{Expsetup}. Here, the fluorescence collected after the dichroic mirror (DM) is split by a beam splitter (BS) into two detection paths, coupled through \(L_6\) and \(L_7\) to two fibers, detected by two SPADs, and subsequently recorded using a time tagger for measuring \(g^{(2)}(\tau)\).}
\label{exp_g2}
\end{figure}

Fig.~\ref{exp_g2} shows the purposed modified experimental setup used for the correlation measurements. While the excitation path remains same, the collection path is modified after the fluorescence passes through the dichroic mirror. The collected fluorescence is split by a beam splitter into two detection arms and directed through two collection lenses, which focus the light into two optical fibers connected to two SPADs. The outputs from the SPADs are then sent to a time tagger, enabling the measurement of the second-order correlation function \(g^{(2)}(0)\) for each measurement.

\begin{figure}[ht]
\centering
\includegraphics[scale=0.5]{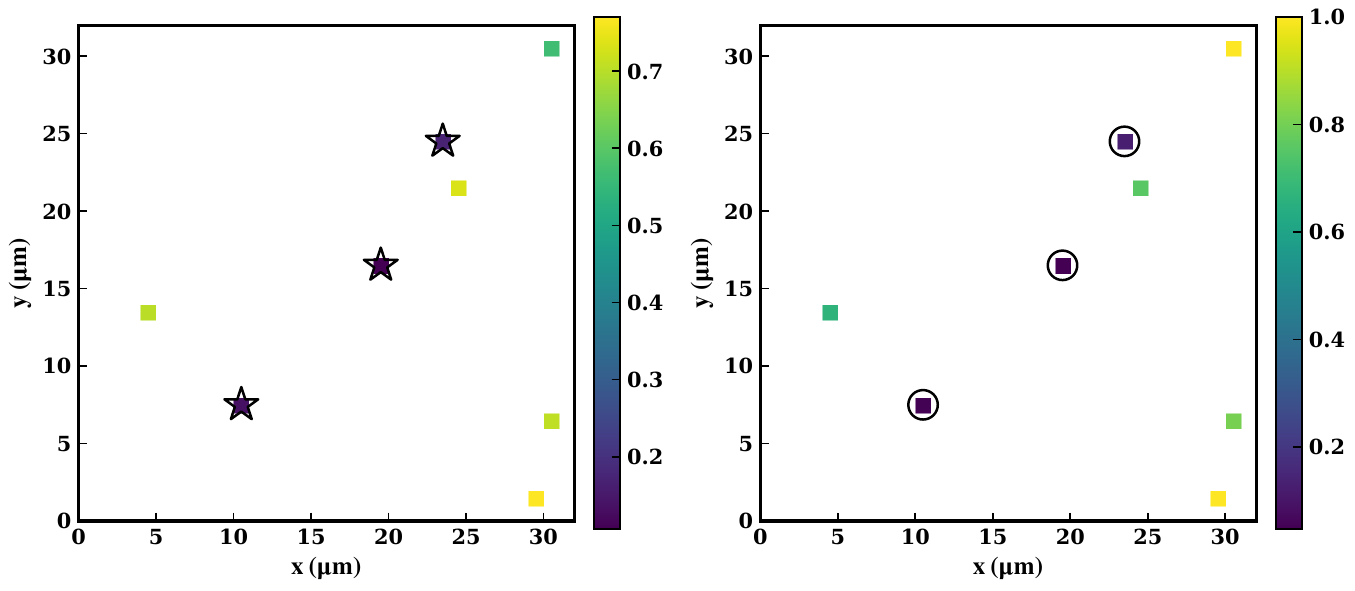}
\caption{Reconstruction of the spatial $g^{(2)}(0)$ map. (a) Direct measurement of $g^{(2)}(0)$ at each raster-scanned position. The star markers indicate locations where $g^{(2)}(0) < 0.5$, identifying single-photon emitters. (b) Reconstruction of the $g^{(2)}(0)$ map using compressive sensing with only $20\%$ of the measurements. The circles indicate positions where the reconstructed $g^{(2)}(0)$ value is below 0.5, demonstrating successful identification of single-photon emitters with significantly reduced data acquisition.}
\label{simg2}
\end{figure}

The reconstructed \(g^{(2)}(0)\) maps are shown in Fig.~\ref{simg2}. Figure~\ref{simg2}(a) shows the \(g^{(2)}(0)\) values obtained from direct raster-scanned correlation measurements. Locations with \(g^{(2)}(0) < 0.5\), corresponding to single-photon emitters, are marked with star symbols. Using the compressive sensing framework described above, the spatial \(g^{(2)}(0)\) map is reconstructed from only \(20\%\) of the measurements, as shown in Fig.~\ref{simg2}(b). The circular markers indicate pixels for which the reconstructed \(g^{(2)}(0)\) value is below \(0.5\), demonstrating that the compressive reconstruction correctly identifies the emitter locations despite substantial undersampling. These results indicate that the proposed method can recover single-photon emitter locations accurately while significantly reducing the number of required measurements.

\section{Conclusion}

In this work, we have demonstrated a compressive sensing imaging framework in which conventional raster scanning is replaced by spatially structured wide-field illumination generated using a digital micromirror device. A sequence of random illumination patterns is applied to the sample, and the resulting collective fluorescence signals are used to reconstruct the image from a reduced number of measurements. The reconstruction is performed using a gradient projection for sparse reconstruction algorithm, exploiting the sparsity of the image in the spatial domain. Using NV centers in diamond as a representative quantum emitter platform, we achieve reliable image reconstruction using only approximately $20\%$ of the measurements required for conventional raster-scanned imaging. Quantitative comparison with raster-scanned reference images using peak signal-to-noise ratio and normalized correlation metrics confirms the accuracy of the reconstructed images despite strong undersampling. These results validate the proposed approach as an efficient and general imaging strategy for sparse quantum emitter systems. In addition to intensity imaging, we extend this framework to reconstruct spatial maps of the second-order correlation function $g^{(2)}(0)$ from compressive measurements through numerical simulations. This enables identification of single-photon emitters via antibunching signatures under significant undersampling, demonstrating the potential of compressive sensing for correlation-based quantum imaging.

The presented method is general and does not rely on emitter-specific properties beyond spatial sparsity, making it directly applicable to a broad class of quantum emitter systems and fluorescence imaging scenarios. By reducing acquisition time and measurement overhead, this approach is particularly advantageous for photon-limited samples, large fields of view, and sparse emitter distributions.

Looking ahead, this framework can be extended through the use of adaptive or learned illumination patterns to further improve reconstruction efficiency. Furthermore, the integration of compressive sensing with correlation-based measurements, such as $g^{(2)}(0)$, opens avenues for simultaneous recovery of spatial and statistical information from quantum emitters. More broadly, the approach can be readily adapted to other quantum and classical imaging platforms where rapid and photon-efficient data acquisition is essential.

\bibliography{ref}

\end{document}